\documentclass{vgtc}                          %
\ifpdf%
  \pdfoutput=1\relax                   %
  \usepackage{graphicx}                %
  \DeclareGraphicsExtensions{.pdf,.png,.jpg,.jpeg} %
\else%
  \usepackage{graphicx}                %
  \DeclareGraphicsExtensions{.eps}     %
\fi%

\graphicspath{{figures/}{pictures/}{images/}{./}} %

\usepackage{microtype}                 %
\PassOptionsToPackage{warn}{textcomp}  %
\usepackage{textcomp}                  %
\usepackage{mathptmx}                  %
\usepackage{times}                     %
\usepackage{cite}                      %
\usepackage{tabu}                      %
\usepackage{booktabs}                  %

\usepackage{xspace}
\usepackage{enumitem}

\usepackage{xcolor}
\onlineid{0}

\vgtccategory{Research}

\vgtcinsertpkg

\newcommand{\ag}[0]{auto-grader\xspace{}}

\definecolor{amber}{rgb}{1.0, 0.75, 0.0}
\definecolor{awesome}{rgb}{1.0, 0.13, 0.32}
\definecolor{bronze}{rgb}{0.8, 0.5, 0.2}
\definecolor{indigo}{rgb}{0.0, 0.25, 0.42}
\definecolor{heliotrope}{rgb}{0.87, 0.45, 1.0}
\definecolor{forestgreen}{rgb}{0.13, 0.55, 0.13}
\definecolor{ginger}{rgb}{0.69, 0.4, 0.0}
\definecolor{jade}{rgb}{0.0, 0.66, 0.42}
\definecolor{mediumslateblue}{rgb}{0.48, 0.41, 0.93}
\definecolor{mint}{rgb}{0.24, 0.71, 0.54}
\definecolor{mulberry}{rgb}{0.77, 0.29, 0.55}
\definecolor{hotpink}{RGB}{255, 83, 115}
\definecolor{aqua}{RGB}{87, 180, 181}
\definecolor{linkcolor}{RGB}{6,125,233}

\newcommand{\hide}[1]{}

\definecolor{purple}{RGB}{173, 99, 252} %
\definecolor{orange}{RGB}{245,142,20} %

\title{Towards Automatic Grading of D3.js Visualizations}

\author{Matthew Hull\thanks{e-mail: matthewhull@gatech.edu}\\ %
        \scriptsize Georgia Institute of Technology %
\and Connor Guerin\thanks{e-mail: cguerin6@gatech.edu}\\ %
     \scriptsize Georgia Institute of Technology %
\and Justin Chen \thanks{e-mail: jchen3001@gatech.edu}\\ %
     \scriptsize Georgia Institute of Technology %
\and Susanta Routray \thanks{e-mail: sroutray@gatech.edu}\\ %
     \scriptsize Georgia Institute of Technology %
\and Duen Horng (Polo) Chau \thanks{e-mail: polo@gatech.edu}\\ %
     \scriptsize Georgia Institute of Technology %
     }

\teaser{
  \centering
  \includegraphics[width=\linewidth]{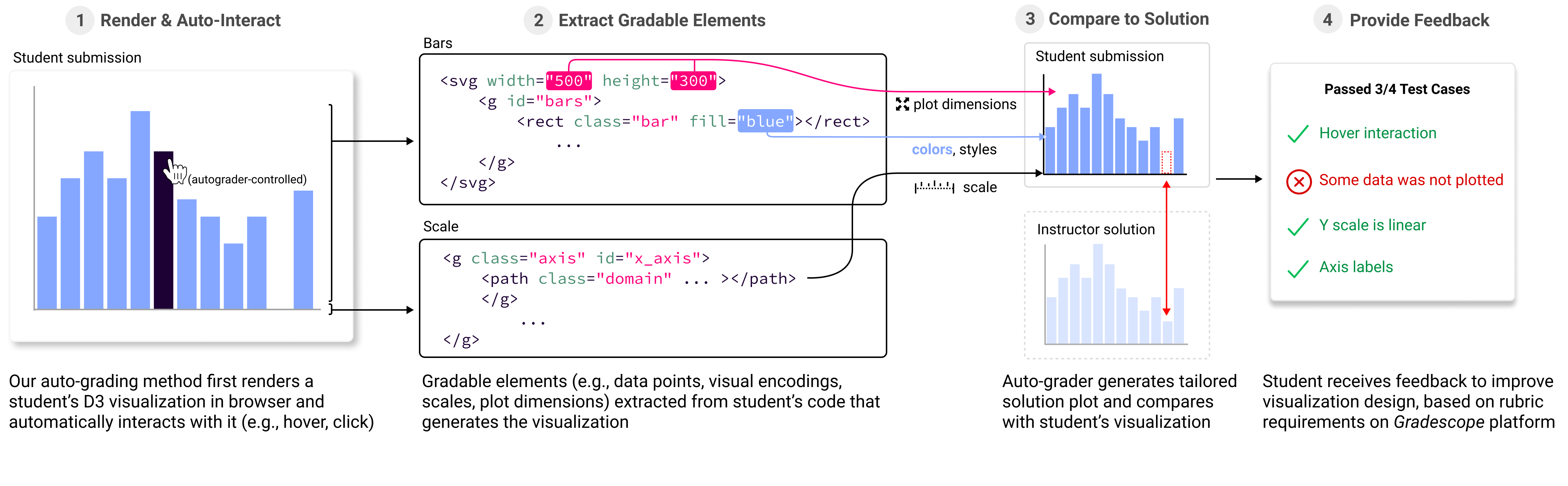}
  \caption{Our auto-grading workflow for D3 visualizations 
  (1) first renders student's plot in a web browser and automatically interacts with it (e.g., hover, click); 
  (2) extracts gradable elements (e.g., data points, visual encodings, plot dimensions) from the plot's backing code; 
  (3) generates a student-specific solution and compares it with student's plot to identify issues; and 
  (4) provides feedback to help student improve their design.}
  \label{fig:teaser}
}

\abstract{
Manually grading D3 data visualizations is a challenging endeavor, and is especially difficult for large classes with hundreds of students.
Grading an interactive visualization requires a combination of interactive, quantitative, and qualitative evaluation that are conventionally done manually and are difficult to scale up as the visualization complexity, data size, and number of students increase.
We present a first-of-its kind automatic grading method for D3 visualizations
that scalably and precisely evaluates the data bindings, visual encodings, interactions, and design specifications used in a visualization.
Our method has shown potential to enhance students' learning experience, enabling them 
to submit their code frequently and receive rapid feedback to better inform iteration and improvement to their code and visualization design. 
Our method
promotes consistent grading and enables instructors to dedicate more focus to 
assist students in gaining visualization knowledge and experience.
We have successfully deployed our method and auto-graded D3 submissions from
more than 1000 undergraduate and graduate students in Georgia Tech's \textit{CSE6242 Data and Visual Analytics} course, and received positive feedback and encouragement for expanding its adoption. 
} %

\CCScatlist{
  \CCScatTwelve{Human-centered computing}{Visu\-al\-iza\-tion}{Visualization design and evaluation methods}{}
}

\begin{document}

\firstsection{Introduction}

\maketitle

Manually grading D3 \cite{2011-d3} data visualizations is a challenging endeavor,
and is especially difficult for large classes with hundreds of students.
Even for a ``simpler'' interactive bar plot visualization, 
manually grading it usually requires a combination of
interactive, quantitative, and qualitative evaluation,  
with tasks such as 
verifying the correctness of data bindings and visual encodings,
testing interactive elements and actions work as intended (e.g., whether tooltip displays upon mouse-hover), 
and determining if a visualization's overall layout and spacing is appropriate.
However, such ``simple''
evaluation tasks quickly become daunting
as the visualization complexity and data size increase. 
For a large class with hundreds or a thousand students,
such manual grading becomes extremely tedious.
Furthermore, manual grading presents the logistic challenge of requiring congruence between student and grader environments. This necessitates the recreation of the student's environment or in some cases, modification of student code by the grader to run the visualization, resulting in a cumbersome grading process, especially for large classes with many students.
To address the common challenges in evaluating D3 visualizations, our ongoing work makes the following contributions:

\begin{enumerate}[itemsep=2mm, topsep=2mm, parsep=1mm, leftmargin=5mm] 

\item{\textbf{We present a first-of-its-kind automatic grading approach for D3 visualizations}} 
that scalably and precisely evaluates the data bindings, visual encodings, interactions, and design specifications used in a visualization.
Our method avoids a rigid ``one-size-fits-all'' approach to grading, by offering a novel flexible way that provides students with tailored feedback while supporting their design freedom in developing visualizations. 
For students' visualizations to be scalably auto-graded using our approach, 
the only requirement is to assign identifiers to a small number of high-level structures of a visualization so that ``gradable'' elements may be extracted for grading. 
Our approach is inspired by Harper and Agrawala's D3 deconstruction approach \cite{harper2014deconstructing}. 
For example, assigning the ``bar'' CSS class name to the group of bars in bar plot (see \autoref{fig:teaser}.2) enables all the bars and their visual properties to be precisely extracted.
Our approach scalably and systematically accomplishes important grading tasks that are otherwise tedious, 
such as verifying whether all data are plotted,  or correct scale types are used.
Our approach provides ample room for students to exercise their design freedom.
For example,
as illustrated in \autoref{fig:teaser}.3, when evaluating a student's bar plot,
the auto-grader generates a \textit{solution} plot specifically for that student (e.g., using plot dimensions, color schemes chosen by student),
then compares the solution plot with the student's plot via instructor-supplied \textit{test cases} that help students identify issues.
Students receive textual feedback on test cases that they pass or fail, and visual feedback in the form of a downloadable image containing their rendered visualization produced during auto-grading (\autoref{fig:teaser}.4).
Students can run the \ag{} as often as they want before the assignment due date,
students' learning experience is enhanced through such frequent feedback while they iteratively improve their work while gaining clarity in how their work is graded.

\item{\textbf{Grading Interactivity.}}
D3 visualizations may be designed to incorporate a myriad of interactive features, ranging from simple actions such as \textit{hover} and \textit{click}, to more complex actions such as filtering a dataset,  drag-drop, or displaying a subplot.
We address this challenge to grade interactivity by utilizing Selenium's \cite{Selenium} browser automation package to compose rich interaction sequences, known as \textit{Action Chains}, that facilitate the evaluation of common user interaction. 
As a result, our \ag{} can efficiently and automatically interact with and grade interactions and ``hidden'' visualization elements that display after a complex sequence of interactions. 
Without our approach, such grading would need to be tediously performed for every student's submission. 

\item{\textbf{Large Scale D3 Auto-Grading: Early Usage.}}
To demonstrate the feasibility and scalability of our approach, we have successfully deployed our \ag{} on the Gradescope\cite{Gradescope} platform, and auto-graded D3 visualization submissions from over 1000 undergraduate and graduate students in Georgia Tech's \textit{CSE6242 Data and Visual Analytics}\footnote{\url{https://poloclub.github.io/\#cse6242}} course during the Spring 2021 semester.
We have enjoyed positive feedback from both students and members of the instruction team.
Excited with this early success, 
we proceeded and 
have completed the work necessary to release the first entirely auto-graded D3 class assignment for the Fall 2021 semester;
the assignment consists of designing and implementing 4 complex interactive data visualizations.

\end{enumerate}

\medskip %

\section{Auto-Grading Impact and Initial Feedback}

\subsection{Student Impact}
    
\noindent
\textbf{Improved Student-Instructor Interaction.}
Since deploying the \ag{}, we observed its positive effects on our \textit{Data and Visual Analytics} course's 1000+ students at Georgia Tech
via student activity on Piazza, the course's discussion forum.
Specifically,
by comparing 
the content of Piazza posts across semesters, 
we found that the \ag{} had greatly reduced the number of question posts related to grading (\textit{``would I lose points for \rule{0.5cm}{0.1mm}?''}), 
acceptable code functions (\textit{``can I use \rule{0.5cm}{0.1mm}?''}),
and visualization styling. 
This reduction was likely due to our approach supporting flexibility in students' design choices and offering immediate feedback as they evolved their visualizations.
We are excited by 
having increased student-instructor interaction on more qualitative topics about gaining knowledge and experience with visualization design and implementation, and less on logistical issues that could distract students.

\smallskip
\noindent
\textbf{Reducing Configuration Errors.} Our approach has dramatically reduced configuration errors that students may experience, such as incorrectly referenced libraries or data files;
before using \ag{}, such errors would prevent the visualization from being rendered or graded, and graders would need to manually resolved them.
The \ag{} re-creates the student's development environment and eliminates potential system configuration mismatch between student and instructor machines.  
Each semester, a student may request that an assignment be regraded if they feel there was an error in the grading process. 
We saw that our auto-graded D3 question received only 1 regrade request, compared to 22 in the previous semester for the same question that was manually graded, 19 of which were due to system configuration issues that could not have occurred in the unified \ag{} environment.

\subsection{Instructor Impact}
Our method has greatly reduced the time and effort required from instructors when evaluating students' work. In previous semesters, it would take each grader hours to assess 100--150 student submissions for an assignment question.  
With the auto-grading method in place, 
not only was the overall grading time and effort reduced, 
but so was the effort needed for processing regrade requests and responding to discussion posts on Piazza.  
Indeed, since students' work was immediately auto-graded as they ran the auto-graders, 
grading was effectively ``done'' as soon as the assignment was due.
In other words,
the instruction team only needed to review a small  number of submissions that received anomalous scores (e.g., a student's final submission received lower scores than earlier submissions).
We found that with this reduction in workload and subsequent increase in availability, we can dedicate more focus to instruction and assisting students.

\section{Conclusion and Future Work}
We believe our auto-grading method is a major first step in scaling up instruction and evaluation of D3 visualizations.
We are excited by our successful deployment and the positive feedback from students and members of the instruction team.
We plan to extend our \ag{}'s  qualitative evaluation ability to consider aspects like color choices, to ensure that the visualization is readable and accessible. 
To further improve students' learning experience, we plan to provide more visual feedback, such as highlighting parts of a visualization that cause an issue, similar to the annotation method used by Hopkins \textit{et al.} \cite{hopkins2020visualint}.
Our auto-grading method is not limited to pure D3 applications --- it is readily extendable to other visualization platforms based on D3, 
which include popular visualization tools such as Vega-Lite\cite{2017-vega-lite}, Observable Notebooks \cite{ObservableNotebook}, and the recently-released Observable Plot \cite{ObservablePlot}.

\bibliographystyle{abbrv-doi}

\bibliography{template}

\end{document}